\documentclass{article}
\usepackage{graphicx}
\usepackage[letterpaper,top=2cm,bottom=2cm,left=3cm,right=3cm,marginparwidth=1.75cm]{geometry}
\usepackage{amsmath,amssymb,amsfonts,hyperref,authblk,booktabs}
\usepackage[numbers,sort&compress]{natbib}
\makeatletter
\def\BibitemShut#1{\unskip\csname bibitem#1\endcsname}
\makeatother

\numberwithin{equation}{section}
\allowdisplaybreaks

\newcommand{\Tr}{\operatorname{Tr}}

\title{Coarse graining without observables:\\ the three-dimensional Ising model in the derivative expansion}
\author[1,2]{Alfio Bonanno}
\affil[1]{\small INAF -- Osservatorio Astrofisico di Catania, Via S. Sofia 78, 95123 Catania, Italy}
\affil[2]{\small INFN, Sezione di Catania, Via S. Sofia 64, 95123 Catania, Italy}
\date{}

\begin{document}
\maketitle

\begin{abstract}
A new approach to the problem of the residual regulator dependence in functional renormalization group calculations
is proposed. The coarse graining implemented by the infrared cutoff is constrained by means of an exact identity
of the flow, which determines the response of the effective average action to a change of the cutoff function. The
cutoff functions for which the violation of this identity is minimal form a Pareto front. For the three-dimensional Ising model at
fourth order of the derivative expansion the critical exponents on the front are
$\eta=0.03435(47)$, $\nu=0.63038(15)$ and $\omega=0.8256(5)$, where the errors represent the dispersion along the
front. The stationarity of the anomalous dimension, which is imposed in the principle of minimal sensitivity, follows
from the identity as a necessary condition, while the other exponents are not stationary. In this approach no observable is needed
in order to select the regulators.
\end{abstract}

\section{Introduction}
\label{sec:intro}

The functional renormalization group (RG) based on the effective average action $\Gamma_k$
\cite{Wetterich:1992yh,Morris:1993qb,Berges:2000ew,Dupuis:2020fhh} is an exact framework, but every application
relies on a truncation of the space of action functionals. A truncated flow depends on the infrared regulator $R_k$,
which in the exact theory drops out of all physical quantities, and a prescription is needed in order to choose it.
The prescription in common use is the principle of minimal sensitivity (PMS) \cite{Stevenson:1981vj,Canet:2002gs}: a
family of regulators with a free parameter is considered, and the parameter is adjusted so that the quantity of
interest is stationary. In the derivative expansion (DE) of the three-dimensional Ising model this procedure, carried
out up to sixth order \cite{Balog:2019rrg,DePolsi:2020pjk}, has produced critical exponents in close agreement with
those of the conformal bootstrap \cite{Kos:2016ysd,Simmons-Duffin:2016wlq}.

The PMS has two features which are unsatisfactory from a conceptual point of view. It takes an observable as its
input, so that the regulator is not determined before the quantity one would like to predict is computed. Moreover it
is applied to each exponent separately \cite{Balog:2019rrg,DePolsi:2020pjk}: the correlation-length exponent $\nu$ is
quoted at the regulator where $\nu$ is stationary and the anomalous dimension $\eta$ at the regulator where $\eta$ is
stationary, so that the exponents quoted at a given order do not belong to one and the same fixed point. The need for
``a more solid theoretical basis'' has been stressed by the authors of Ref.~\cite{DePolsi:2020pjk} themselves. A
criterion which does not refer to an observable has been proposed in Ref.~\cite{Balog:2020fyt}, where the violation of
the Ward identity of special conformal transformations is minimized; it requires conformal invariance of the fixed
point. Functional optimization has been developed as a principle for the choice of the regulator
\cite{Litim:2000ci,Litim:2001up,Pawlowski:2005xe}. Regulator variations and the loss of integrability in truncated
flows have been investigated in Ref.~\cite{Pawlowski:2015mlf}, where optimization criteria based on the regularized
propagator and on the length of the RG trajectories are formulated. A multi-parameter version of the PMS on
$\eta$ has been explored recently in Ref.~\cite{Codello:2026xyz}.

The dependence of $\Gamma_k$ on the regulator is governed by an
exact identity \cite{Pawlowski:2005xe,Pawlowski:2015mlf},
\begin{equation}
\frac{\delta \Gamma_k[\varphi]}{\delta R_k(q)}=\frac12\,\big(\Gamma_k^{(2)}[\varphi]+R_k\big)^{-1}(q,-q)\,,
\label{eq:identity}
\end{equation}
of which the flow equation is the particular case of a change of the scale $k$. Here $\varphi$ is the average field and
$\Gamma_k^{(2)}$ the second functional derivative of $\Gamma_k$. In this work a criterion for the selection of the regulator is formulated
directly from the violation of (\ref{eq:identity}) at a truncated fixed point. The response of the fixed point to a
change of the regulator, obtained by differentiating the fixed-point equations with respect to a parameter of the
regulator, is compared with the response prescribed by (\ref{eq:identity}). Their difference, which will be called
the \emph{defect}, vanishes in the exact theory and measures the violation of the identity within the truncation. The regulator is then selected by requiring that the defect be as small as possible. Since a
truncation at order $\partial^2$ contains two functions, the potential and the wave-function renormalization, there
are two defects, and the regulators are not ranked by a combination of them. A regulator is discarded only if
another regulator has both of them smaller. What is left is a one-parameter set of regulators, a Pareto front, on which the critical
exponents are read off. The construction is carried out at second and at fourth order of the derivative expansion,
and the critical exponents do not enter the selection.

For a rational family of regulators with an amplitude and a sharpness the two
defects constrain the amplitude and restrict the sharpness to an interval. The regulators in common use, at the amplitudes selected by the PMS, turn out to be dominated. The
amplitude selected by the identity lies within about one percent of the amplitude at which $\eta$ is stationary,
and it is shown in Sec.~\ref{sec:pms} that a vanishing defect implies the stationarity of $\eta$, while no analogous
statement holds for $\nu$ or for the correction-to-scaling exponent $\omega$. In this sense the identity explains the success of the PMS when it is applied to $\eta$ and
removes the motivation for applying it exponent by exponent.

Besides its conceptual interest, a criterion of this kind is needed in those applications of the functional RG in
which no suitable observable is available as a target of the optimization and the regulator depends on the fields, as in quantum gravity
or in nonlinear sigma models \cite{Bonanno:2026bilocal}. The Ising model serves here as the benchmark in which the
outcome can be compared with independent results.

\section{The regulator identity at a fixed point}
\label{sec:defect}

\subsection{Flow equations}

The effective average action of a scalar field $\varphi$ with $Z_2$ symmetry in $d$ Euclidean dimensions obeys the flow
equation \cite{Wetterich:1992yh,Morris:1993qb}
\begin{equation}
\partial_t\Gamma_k[\varphi]=\frac12\,\Tr\Big[\partial_tR_k(q^2)\,G_k(q^2,\varphi)\Big]\,,\qquad t=\ln k\,,
\label{eq:flow}
\end{equation}
where $G_k=(\Gamma_k^{(2)}+R_k)^{-1}$ is the regularized propagator. At the first non-trivial order of the derivative
expansion the effective average action is
\begin{equation}
\Gamma_k[\varphi]=\int d^dx\,\Big[V_k(\varphi)+\frac12\,Z_k(\varphi)\,(\partial\varphi)^2\Big]+O(\partial^4)\,,
\label{eq:ansatz2}
\end{equation}
so that for a constant field
\begin{equation}
G_k(q^2,\varphi)=\frac{1}{Z_k(\varphi)\,q^2+V_k''(\varphi)+R_k(q^2)}\,.
\label{eq:prop}
\end{equation}
Here and below a prime indicates a derivative with respect to the argument. The flow of $V_k$ is obtained
from (\ref{eq:flow}) at constant field, and the flow of $Z_k$ from the term of order $p^2$ of the flow of the
two-point function, with $\int_q\equiv\int d^dq/(2\pi)^d$,
\begin{align}
\partial_t\Gamma_k^{(2)}(p^2)={}&\int_q\Big[Z_k'\,\big(q^2+q\!\cdot\!p+p^2\big)+V_k'''\Big]^2
G_k\big((q+p)^2\big)\,G_k^2(q^2)\,\partial_tR_k(q^2)\nonumber\\
&-\frac12\int_q\Big[Z_k''\,\big(q^2+p^2\big)+V_k''''\Big]\,G_k^2(q^2)\,\partial_tR_k(q^2)\,,
\label{eq:gamma2flow}
\end{align}
after the expansion of $G_k((q+p)^2)$ in powers of $p$ and the angular average. With the derivatives of the
propagator with respect to $q^2$,
\begin{equation}
G_k'=-G_k^2\,\big(Z_k+R_k'\big)\,,\qquad G_k''=2\,G_k^3\,\big(Z_k+R_k'\big)^2-G_k^2\,R_k''\,,
\label{eq:propder}
\end{equation}
the result is
\begin{equation}
(4\pi)^{\frac d2}\,\partial_tV_k=\frac12\,Q_{\frac d2}\big[G_k\,\partial_tR_k\big]
\label{eq:BV}
\end{equation}
and
\begin{align}
(4\pi)^{\frac d2}\,\partial_tZ_k={}&\big(V_k'''\big)^2\Big\{Q_{\frac d2}\big[G_k^2G_k'\,\partial_tR_k\big]
+Q_{\frac d2+1}\big[G_k^2G_k''\,\partial_tR_k\big]\Big\}\nonumber\\
&+Z_k'\,V_k'''\,\Big\{2\,Q_{\frac d2}\big[G_k^3\,\partial_tR_k\big]
+(d+2)\,Q_{\frac d2+1}\big[G_k^2G_k'\,\partial_tR_k\big]
+(d+2)\,Q_{\frac d2+2}\big[G_k^2G_k''\,\partial_tR_k\big]\Big\}\nonumber\\
&+\big(Z_k'\big)^2\Big\{\frac{2d+1}{2}\,Q_{\frac d2+1}\big[G_k^3\,\partial_tR_k\big]
+\frac{(d+2)(d+4)}{4}\,Q_{\frac d2+2}\big[G_k^2G_k'\,\partial_tR_k\big]\nonumber\\
&\qquad\qquad+\frac{(d+2)(d+4)}{4}\,Q_{\frac d2+3}\big[G_k^2G_k''\,\partial_tR_k\big]\Big\}
-\frac12\,Z_k''\,Q_{\frac d2}\big[G_k^2\,\partial_tR_k\big]\,,
\label{eq:BZ}
\end{align}
where the $Q$-functionals of the variable $z=q^2$ are defined as
\begin{equation}
Q_n[f]\equiv\frac{1}{\Gamma(n)}\int_0^\infty dz\,z^{n-1}f(z)\,.
\label{eq:Qfun}
\end{equation}
These relations are valid for an arbitrary regulator and in any dimension \cite{Bonanno:2022edf}. The regulator is
written as
\begin{equation}
R_k(q^2)=\bar Z_k\,k^2\,r(y)\,,\qquad y=q^2/k^2\,,
\label{eq:regulator}
\end{equation}
with a dimensionless profile $r(y)$ and a field-independent normalization $\bar Z_k$, which defines the anomalous
dimension $\eta=-\partial_t\ln\bar Z_k$, so that
\begin{equation}
\partial_tR_k=\bar Z_k\,k^2\,\dot r(y)\,,\qquad \dot r(y)\equiv(2-\eta)\,r(y)-2\,y\,r'(y)\,.
\label{eq:rdot}
\end{equation}

A fixed point is a $k$-independent solution of the flow in terms of the rescaled variables
\cite{Berges:2000ew,Balog:2019rrg,DePolsi:2020pjk}
\begin{equation}
\varphi=\bar Z_k^{-1/2}\,k^{\frac{d-2}{2}}\,x\,,\qquad V_k(\varphi)=k^d\,v(x)\,,\qquad Z_k(\varphi)=\bar Z_k\,z(x)\,.
\label{eq:rescaling}
\end{equation}
At a fixed point $\bar Z_k\propto k^{-\eta}$, so that the field is rescaled with $k^{(d-2+\eta)/2}$ and $Z_k$ with
$k^{-\eta}$, and the solution obeys
\begin{equation}
0=-d\,v+D_\varphi\,x\,v'+b_V[\dot r]\,,\qquad
0=\eta\,z+D_\varphi\,x\,z'+b_Z[\dot r]\,,\qquad D_\varphi=\frac{d-2+\eta}{2}\,.
\label{eq:fp2}
\end{equation}
Here $b_V$ and $b_Z$ are the loop terms of the two sectors. For a function $h(y)$,
\begin{equation}
b_V[h]\equiv\frac12\,Q_{\frac d2}\big[g\,h\big]\,,\qquad g(y)=\frac{1}{z\,y+v''+r(y)}\,,
\label{eq:gdimless}
\end{equation}
and $b_Z[h]$ is the right-hand side of (\ref{eq:BZ}) with $V_k$, $Z_k$, $G_k$ and $\partial_tR_k$ replaced by $v$,
$z$, $g$ and $h$, where the $Q$-functionals act on functions of $y$ and the primes on $g$ are derivatives with respect
to $y$. The factor $(4\pi)^{-d/2}$ has been absorbed in a rescaling of $x$ and $v$. Both are linear in the insertion
$h$, which in (\ref{eq:fp2}) is $\dot r$. The normalization of $z$ is fixed
at the minimum $x_0$ of the potential,
\begin{equation}
v'(x_0)=0\,,\qquad z(x_0)=1\,,
\label{eq:norm}
\end{equation}
which amounts to the choice $\bar Z_k=Z_k(\varphi_0)$, and these two conditions determine $x_0$ and $\eta$.

\subsection{Response to a change of the regulator}

Let now the profile depend on a parameter $\theta$, with $\delta r(y)\equiv\partial_\theta r(y)$. The solution
$f^*=(v^*,z^*)$ of (\ref{eq:fp2}) depends on $\theta$, and its derivative $\partial_\theta f^*$ at fixed $x$ is the
response of the truncated fixed point; it is obtained from the linearized fixed-point equations. On the
other hand, Eq.~(\ref{eq:identity}) states that a change $\delta R_k$ of the regulator changes $\Gamma_k$ by
$\frac12\Tr[\delta R_k\,G_k]$, which is the right-hand side of (\ref{eq:flow}) with $\partial_tR_k$ replaced by
$\delta R_k$. Its projection on the ansatz (\ref{eq:ansatz2}) is therefore given by the same functionals $b_V$ and
$b_Z$ with the insertion $\delta r$ in the place of $\dot r$. Since the regulator contains the
normalization $\bar Z_k$, which may also change with $\theta$, the response prescribed by the identity is
$b_f[\delta r]+\gamma\,N_f$, $f=v,z$, with
\begin{equation}
N_v=b_V[r]-\frac12\,x\,v'\,,\qquad N_z=b_Z[r]-z-\frac12\,x\,z'\,.
\label{eq:exactside}
\end{equation}
The function $N_f$ accounts for the variation of the normalization in the regulator and in the rescaled variables;
it is minus the derivative of the equations (\ref{eq:fp2}) with respect to $\eta$ at fixed $v$ and $z$. In the exact
theory $\gamma=\partial_\theta\ln\bar Z_k$. The fixed-point equations determine $\eta$ and not the value of
$\bar Z_k$, so that $\gamma$ is not provided by them, and it is kept as a parameter. The defect is the difference of
the two responses,
\begin{equation}
D_f=\partial_\theta f^*-\big(b_f[\delta r]+\gamma\,N_f\big)\,.
\label{eq:defectdef}
\end{equation}
For the exact fixed point it vanishes, with $\gamma=\partial_\theta\ln\bar Z_k$. In a truncation it vanishes for every
profile, with $\gamma=-\eta$, if the variation is a change of the scale $k$, which corresponds to
$\delta r=2\,(r-y\,r')$.

In the following the parameter $\theta$ is the logarithm of the overall amplitude of the regulator. Writing
$r(y)=\alpha\,\hat r(y)$, with $\hat r$ a fixed function, and $\theta=\ln\alpha$, one has
\begin{equation}
\delta r(y)=\alpha\,\partial_\alpha r(y)=r(y)\,,
\label{eq:amplitude}
\end{equation}
so that the regulator is multiplied by the same factor at all momenta and its shape is not modified.

The loop terms $b_V$ and $b_Z$ contain $v$ only through its derivatives, so that the
field-independent part of $v$ does not act on the rest of the solution: it is obtained afterwards from the first of
Eqs.~(\ref{eq:fp2}) at $x=0$, $v(0)=b_V[\dot r](0)/d$. Since $v$ is even, $v'$ vanishes linearly at the origin, and the potential is represented by
the even function
\begin{equation}
u(x)\equiv\frac{v'(x)}{x}\,,
\label{eq:udef}
\end{equation}
the derivative of $v$ with respect to $x^2/2$, whose value at the origin is the mass term $v''(0)$. Since the derivative with respect to
$\theta$ is taken at fixed $x$, it commutes with this operation, and the
functions of this sector are obtained from those of $v$ as $D_u=D_v'/x$ and $N_u=N_v'/x$.

The defect is evaluated on a finite interval of the field. The identity (\ref{eq:identity}) is a statement about the
fluctuations, which enter Eqs.~(\ref{eq:fp2}) through the loop terms $b_V$ and $b_Z$. These terms decrease with $x$,
because the propagator is suppressed by the growing mass $v''(x)$, and at large field the fixed-point equations
reduce to their scaling terms, namely to the tree level. In that region the solution is a power law fixed by the
dimension of the field, both responses in (\ref{eq:defectdef}) follow from it, and their difference no longer probes
the loop terms. The comparison is therefore restricted to the interval
$[0,x^*]$ in which the fluctuations are active. Let
\begin{equation}
\ell(x)\equiv\frac1x\,\partial_x\,b_V[\dot r]
\label{eq:loopterm}
\end{equation}
be the loop term in the equation for $u$. The end point $x^*$ is defined as the field at which this term has
decreased by one order of magnitude with respect to its value at the origin,
\begin{equation}
\big|\ell(x^*)\big|=\frac{1}{10}\,\big|\ell(0)\big|\,.
\label{eq:window}
\end{equation}
For the regulators considered below this gives $x^*\simeq2.1\,x_0$, and at $x^*$ the loop term is about one percent
of the scaling terms in the equation for $u$. The interval thus contains the origin, the minimum of the potential and
the region around it where the fixed point is determined by the fluctuations. The dependence of the selection on
the threshold is examined at the end of Sec.~\ref{sec:de4}; the selected amplitude changes by one percent at most when
the threshold is halved or doubled.

On the interval $[0,x^*]$ the scalar product and the norm
\begin{equation}
\langle f,h\rangle\equiv\int_0^{x^*}dx\,x\,f(x)\,h(x)\,,\qquad \lVert f\rVert^2\equiv\langle f,f\rangle
\label{eq:scalar}
\end{equation}
are introduced. Because of the $Z_2$ symmetry $u$ and $z$ are functions of $x^2/2$, and since $dx\,x=d(x^2/2)$ this
is the ordinary $L^2$ scalar product in that variable. Each sector is measured in units of the response of the fixed
point to the change of the regulator, and the two relative defects are
\begin{equation}
\Delta_V=\frac{\lVert D_u\rVert}{\lVert\partial_\theta u^*\rVert}\,,\qquad
\Delta_Z=\frac{\lVert D_z\rVert}{\lVert\partial_\theta z^*\rVert}\,,
\label{eq:objectives}
\end{equation}
the violations of the identity in the potential and in the wave-function sector relative to the response of the fixed
point. At each regulator $\gamma$ is chosen so as to minimize $\Delta_V^2+\Delta_Z^2$, which removes from the defect
its component along $N$; the explicit expression is Eq.~(\ref{eq:gammafix}) of Appendix~\ref{app:norm}. The value of $\gamma$ for which
the response prescribed by the identity respects the normalization condition (\ref{eq:norm}) is in general
different, and it depends on the point at which the field is normalized. The two resulting defects are used
separately in the selection.

A regulator $A$ is said to be dominated by a regulator $B$ if $\Delta_V^B\le\Delta_V^A$ and $\Delta_Z^B\le\Delta_Z^A$,
one of the two inequalities being strict. The regulators of a given family which are not dominated by any other
member form the front of the family.

\section{The cyclotomic regulator}
\label{sec:cyclotomic}

The regulators considered in this paper are
\begin{equation}
r(y)=\frac{\alpha}{1+y+y^2+\dots+y^m}=\alpha\,\frac{1-y}{1-y^{m+1}}\,,
\label{eq:cyclotomic}
\end{equation}
with two parameters, the amplitude $\alpha=r(0)$ and the sharpness $m$. For integer $m$ the denominator is the product of the cyclotomic
polynomials $\Phi_n(y)$ with $n>1$ dividing $m+1$, and it coincides with $\Phi_{m+1}(y)$ when $m+1$ is prime; its
roots are the $(m+1)$-th roots of unity different from one. For this reason the profile will be called the
\emph{cyclotomic} regulator; the second form in (\ref{eq:cyclotomic}) defines it for real $m$. It is a particular choice
of coefficients in the algebraic class of Ref.~\cite{Floerchinger:2011sc}. For $\alpha=1$ the inverse free propagator
is flat at small momenta up to a high order,
\begin{equation}
y+r(y)=\frac{1-y^{m+2}}{1-y^{m+1}}=1+y^{m+1}+O\big(y^{m+2}\big)\,,
\label{eq:flat}
\end{equation}
and for $m\to\infty$ the profile tends to the optimized regulator $r=(1-y)\,\theta(1-y)$ of
Refs.~\cite{Litim:2000ci,Litim:2001up}, where $\theta(1-y)$ is the step function. The sharpness therefore interpolates between a smooth cutoff with a power-law
tail, $r\sim\alpha\,y^{-m}$, and the optimized one, while keeping the profile analytic on the positive real axis.

\begin{figure}[t]
\centering
\includegraphics[width=0.9\textwidth]{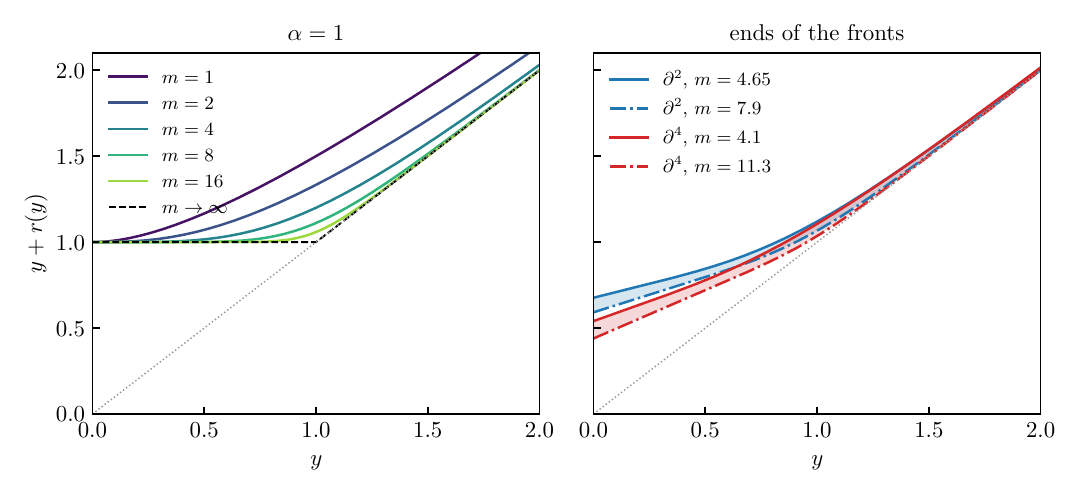}
\caption{The inverse free propagator $y+r(y)$ for the cyclotomic regulator (\ref{eq:cyclotomic}). Left: $\alpha=1$ and
several values of the sharpness; the dashed line is the optimized regulator. Right: the regulators at the two ends of
the fronts of Sects.~\ref{sec:de2} and \ref{sec:de4}. The dotted line is $y$.}
\label{fig:cyclotomic}
\end{figure}

For $\alpha=1$ the plateau (\ref{eq:flat}) extends towards $y=1$ as $m$ grows (Fig.~\ref{fig:cyclotomic}, left
panel). The regulators selected in the following sections have an amplitude smaller than one, so that at small
momenta $y+r$ grows almost linearly from the value $\alpha$, with slope $1-\alpha$, and joins the free behavior
around $y\simeq1$ (right panel).

A practical advantage of (\ref{eq:cyclotomic}) is that for integer $m$ the momentum integrals can be performed in
closed form. With $E(y)=1+y+\dots+y^m$ the propagator (\ref{eq:gdimless}) is $g=E/\mathcal P$, where
\begin{equation}
\mathcal P(y)=\big(z\,y+v''\big)\,E(y)+\alpha
\label{eq:Qpoly}
\end{equation}
is a polynomial of degree $m+1$ with roots $y_j$. In $d=3$ the $Q$-functionals in (\ref{eq:BZ}), and those in the
derivative of (\ref{eq:BV}) with respect to the field, which is what enters the equation for $u$, reduce to integrals
of the form
\begin{equation}
\int_0^\infty dy\,\sqrt y\,\frac{\mathcal N(y)}{\mathcal P(y)^n}=-\pi\sum_{j=1}^{m+1}\,\underset{y=y_j}{\mathrm{Res}}\,
\frac{\mathcal N(y)\,\sqrt{-y}}{\mathcal P(y)^n}\,,
\label{eq:residues}
\end{equation}
with a polynomial $\mathcal N$ and $n=2,\dots,5$, the square root being positive on the negative real axis. The same property allows one to evaluate in closed form the spectral sums on
a sphere. In the numerical work reported below the integrals have been computed by quadrature, and
(\ref{eq:residues}) has served as an independent check. Explicit expressions, and the position of the poles for one of the selected
regulators, are given in Appendix~\ref{app:closed}.

\section{Selection at order $\partial^2$}
\label{sec:de2}

The fixed-point equations (\ref{eq:fp2}) have been solved on the half-line $x\in[0,\infty)$ by a pseudospectral
method with two domains \cite{Borchardt:2015rxa}, in which the functions divided by their leading large-field behaviour
are expanded in rational Chebyshev functions in the outer domain, so that no boundary condition is imposed at either
end. The exponents $\nu$ and $\omega$ are obtained from the linearized flow around the fixed point. For the exponential
regulator $r=\alpha\,y/(e^y-1)$ the known values of the order $\partial^2$ are recovered \cite{Canet:2002gs}.

The candidates are the cyclotomic regulators with sharpness $m$ from $3.5$ to $12$ in steps of $0.25$ and, for each
$m$, with the same set of amplitudes, $\alpha$ from $0.40$ to $0.95$ in steps of $0.005$. For a given $m$ a regulator
whose amplitude lies outside the interval between the minimum of $\Delta_V$ and the minimum of $\Delta_Z$ is dominated
by a regulator with the same $m$. The front has therefore been obtained from a second set of candidates, with $m$
from $3.6$ to $10$ in steps of $0.025$: for each $m$ the two minima are located by a minimization with respect to
$\alpha$, and twelve amplitudes are taken between them, so that no exponent is used in placing the candidates.

\begin{figure}[t]
\centering
\includegraphics[width=\textwidth]{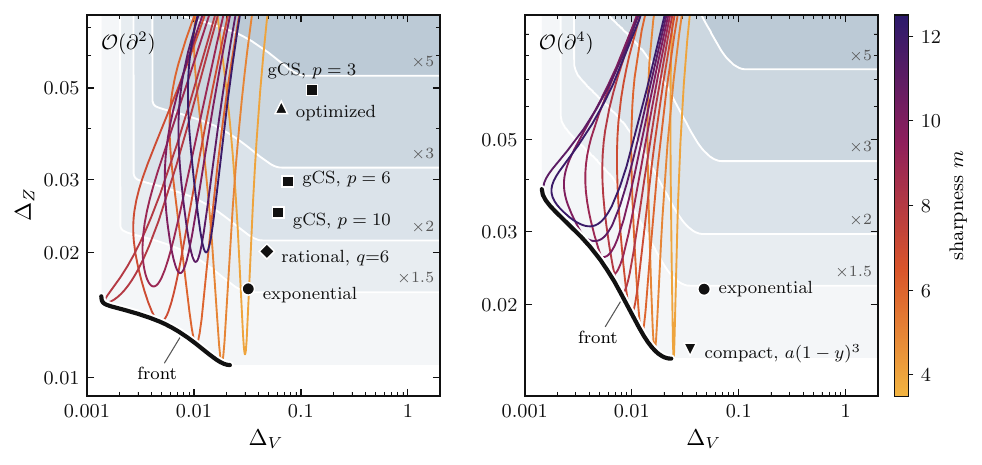}
\caption{The two relative defects $\Delta_V$ and $\Delta_Z$ of Eq.~(\ref{eq:objectives}) at order $\partial^2$ (left) and
$\partial^4$ (right). Each thin curve is a cyclotomic regulator of given integer sharpness $m$ (colour), the amplitude
$\alpha$ varying along the curve. The thick line is the front. The symbols are other regulators at the value of their parameter at
which $\eta$ is stationary. The shading gives the factor by which both defects of a regulator exceed those of some
regulator of the front.}
\label{fig:front}
\end{figure}

The result is shown in the left panel of Fig.~\ref{fig:front}. For a given $m$ the curve described in the plane
$(\Delta_V,\Delta_Z)$ when $\alpha$ is varied has a sharp tip: $\Delta_Z$ has a single, V-shaped minimum as a function of the
amplitude, and it increases steeply on both sides of it. The amplitude is therefore narrowly constrained. When $m$ is increased the
tip moves towards smaller $\Delta_V$ and larger $\Delta_Z$, the minimum of $\Delta_V$ being deepest at $m\simeq7.8$, and
beyond this value both defects grow. The front follows closely the envelope of the tips. It consists of the regulators with
\begin{equation}
4.65\le m\le 7.9\,,\qquad 0.592\le\alpha\le0.676\,,
\label{eq:front2}
\end{equation}
and both ends lie well inside the range which has been scanned. Some of its points are listed in
Table~\ref{tab:front2}.

\begin{table}[t]
\centering
\begin{tabular}{ccccccc}
\toprule
$m$ & $\alpha$ & $\Delta_V$ & $\Delta_Z$ & $\eta$ & $\nu$ & $\omega$ \\
\midrule
7.88 & 0.5919 & 0.0014 & 0.0152 & 0.043629 & 0.628502 & 0.84810 \\
7    & 0.6057 & 0.0046 & 0.0137 & 0.043679 & 0.628481 & 0.84806 \\
6    & 0.6274 & 0.0105 & 0.0120 & 0.043774 & 0.628439 & 0.84804 \\
5    & 0.6607 & 0.0182 & 0.0108 & 0.043940 & 0.628361 & 0.84806 \\
4.65 & 0.6764 & 0.0217 & 0.0107 & 0.044024 & 0.628319 & 0.84809 \\
\bottomrule
\end{tabular}
\caption{Regulators on the front at order $\partial^2$, with their defects and critical exponents. The first and the
last line are at the two ends of the front.}
\label{tab:front2}
\end{table}

All the exponents are computed at each point of the front, i.e.\ at one fixed point, and their ranges over the front
are
\begin{equation}
\eta=0.04383(20)\,,\qquad \nu=0.62841(9)\,,\qquad \omega=0.84807(3)\,,
\label{eq:exp2}
\end{equation}
where the number in brackets is half of the range. It represents the dependence on the regulator which the identity
does not resolve at this order. It is not an estimate of the error of the truncation, which is much larger: the value
of $\eta$ obtained from the conformal bootstrap is $0.036298$ \cite{Kos:2016ysd}.

The regulators which are normally used, each at the value of its parameter at which $\eta$ is stationary, are compared with the
front in Fig.~\ref{fig:front} and in Table~\ref{tab:standard}, where the generalized
Callan-Symanzik regulator is $r=(b/(y+b))^p$, with the width $b$ as the parameter, and the amplitude of the optimized regulator
$r=\alpha(1-y)\theta(1-y)$ is allowed to vary. All of them are dominated. With respect to the exponential regulator,
for $m=7$ the defect in the wave-function sector is about $16\%$ smaller and the one in the potential sector seven
times smaller. The optimized regulator at $\alpha=1$, which is singled out in the local potential approximation
\cite{Litim:2000ci,Pawlowski:2005xe}, is far from the front at this order, with $(\Delta_V,\Delta_Z)=(0.24,0.41)$.

\begin{table}[t]
\centering
\begin{tabular}{lccc}
\toprule
regulator & $\Delta_V$ & $\Delta_Z$ & $\eta$ \\
\midrule
cyclotomic, $m=7$ (on the front) & 0.0046 & 0.0137 & 0.043679 \\
exponential & 0.0322 & 0.0163 & 0.044263 \\
generalized Callan-Symanzik, $p=6$ & 0.0753 & 0.0297 & 0.045380 \\
generalized Callan-Symanzik, $p=3$ & 0.1267 & 0.0494 & 0.045969 \\
optimized, $\alpha=0.494$ & 0.0657 & 0.0449 & 0.043987 \\
\bottomrule
\end{tabular}
\caption{The defects at order $\partial^2$ of some regulators in common use, at the value of their parameter at which $\eta$ is
stationary, compared with a regulator of the front.}
\label{tab:standard}
\end{table}

\section{Selection at order $\partial^4$}
\label{sec:de4}

At the next order of the DE the ansatz contains three more functions,
\begin{equation}
\Gamma_k[\varphi]=\int d^dx\,\Big[V_k+\frac12\,Z_k\,(\partial\varphi)^2
+\frac12\,W^a_k\,(\partial_\mu\partial_\nu\varphi)^2
+\frac12\,W^b_k\,(\partial\varphi)^2\,\partial^2\varphi
+\frac12\,W^c_k\,\big((\partial\varphi)^2\big)^2\Big]\,,
\label{eq:ansatz4}
\end{equation}
all of them depending on $\varphi$, with $W^b_k$ odd. All the contributions of this ansatz to the flows of the five functions
are retained. Denoting by $f=(v,z,w_a,w_b,w_c)$ the rescaled functions of $x$, the fixed-point
equations are
\begin{equation}
0=\lambda_f\,f+D_\varphi\,x\,f'+b_f[\dot r]\,,\qquad
\lambda_f=\big(-d,\;\eta,\;2+\eta,\;2+D_\varphi+\eta,\;d+2\eta\big)\,,
\label{eq:fp4}
\end{equation}
and the propagator is now
\begin{equation}
g(y)=\frac{1}{w_a\,y^2+z\,y+v''+r(y)}\,.
\label{eq:prop4}
\end{equation}
The loop terms $b_f$ have been derived by expanding the flow of the two-, three- and four-point functions to fourth
order in the external momenta; they contain about eight hundred terms and are given in Appendix~\ref{app:flows} and in the ancillary files. They have been
checked against the order $\partial^2$, against each other through the relations between the $n$-point functions, and
against published results: for the exponential regulator the maximum of $\eta$ is $0.034556$, to be compared with
$0.034544$ of Ref.~\cite{Balog:2020fyt}, and for the compact regulator $r=a(1-y)^3\theta(1-y)$ it is $0.035097$, to
be compared with $0.035096$ of Ref.~\cite{Codello:2026xyz}. The defect is defined as in Sec.~\ref{sec:defect}, with
\begin{equation}
N_f=b_f[r]-\frac{\partial\lambda_f}{\partial\eta}\,f-\frac12\,x\,f'\,,
\label{eq:N4}
\end{equation}
which is again minus the derivative of (\ref{eq:fp4}) with respect to $\eta$. The selection is performed with the same two
quantities $\Delta_V$ and $\Delta_Z$ as at order $\partial^2$, the same interval (\ref{eq:window}) and the same variation, so that
the two orders can be compared; the relative defects of $w_a$, $w_b/x$ and $w_c$ in the basis (\ref{eq:ansatz4}), evaluated with the same norm and with the value of $\gamma$ determined by $u$ and $z$ alone, lie between $0.04$ and $0.26$ on the front.

At order $\partial^4$ the function $w_a$ is negative at the fixed point, so that the propagator
(\ref{eq:prop4}) has a pole at a large positive value of $y$, of the order of $150$ for the cyclotomic regulators.
For regulators which do not have compact support the integrals over the half-line are therefore not defined, even
when the pole is numerically invisible, as it is for the exponential regulator. The pole lies at momenta where the DE
does not apply. Here the integrals are cut at $y=60$, on both sides of the identity; the results (\ref{eq:exp4}) do not change
when the cut is moved between $30$ and $100$.

The fixed points have been obtained with finite differences of high order on a uniform grid in the field, without
boundary conditions, and checked with an independent Chebyshev discretization (Appendix~\ref{app:numerics}). The candidates are again the cyclotomic
regulators, with $m$ from $3.5$ to $12$ and, for each $m$, the amplitudes from $0.30$ to $0.75$ in steps of $0.005$,
and the scan has been continued up to $m=14$. The front has been obtained as at order $\partial^2$, with $m$ from
$3.6$ to $13.5$ in steps of $0.05$.

The result is shown in the right panel of Fig.~\ref{fig:front}. The picture is the same as at order $\partial^2$:
for every $m$ the defect $\Delta_Z$ has a single sharp minimum in the amplitude, the minima of $\Delta_V$ and $\Delta_Z$ move in
opposite directions when $m$ is varied, and the front follows the envelope of the tips. It now extends over a larger
interval of the sharpness,
\begin{equation}
4.1\le m\le 11.3\,,\qquad 0.440\le\alpha\le0.541\,,
\label{eq:front4}
\end{equation}
and the selected amplitudes are $20\%$ smaller than at order $\partial^2$: the regulator selected at one order is not
the one selected at the next, where it has $\Delta_Z\simeq0.5$. Some points of the front are given in
Table~\ref{tab:front4}. The critical exponents on the front are
\begin{equation}
\eta=0.03435(47)\,,\qquad \nu=0.63038(15)\,,\qquad \omega=0.8256(5)\,.
\label{eq:exp4}
\end{equation}

\begin{table}[t]
\centering
\begin{tabular}{ccccccc}
\toprule
$m$ & $\alpha$ & $\Delta_V$ & $\Delta_Z$ & $\eta$ & $\nu$ & $\omega$ \\
\midrule
11.3  & 0.4396 & 0.0015 & 0.0367 & 0.033874 & 0.630531 & 0.82597 \\
9     & 0.4633 & 0.0059 & 0.0241 & 0.034337 & 0.630389 & 0.82602 \\
8     & 0.4747 & 0.0078 & 0.0214 & 0.034520 & 0.630335 & 0.82598 \\
7     & 0.4871 & 0.0096 & 0.0193 & 0.034677 & 0.630287 & 0.82588 \\
6     & 0.5021 & 0.0125 & 0.0171 & 0.034788 & 0.630250 & 0.82571 \\
5     & 0.5198 & 0.0164 & 0.0155 & 0.034812 & 0.630234 & 0.82543 \\
4.1   & 0.5407 & 0.0234 & 0.0148 & 0.034681 & 0.630258 & 0.82508 \\
\bottomrule
\end{tabular}
\caption{Regulators on the front at order $\partial^4$, with their defects and critical exponents. The first and the
last line are at the two ends of the front.}
\label{tab:front4}
\end{table}

The two orders are compared in Table~\ref{tab:orders}. For the three exponents the values of the conformal bootstrap
lie between those of the two orders. The interval of $\eta$ over the front is two and a half times larger at order
$\partial^4$ than at order $\partial^2$.

\begin{table}[t]
\centering
\begin{tabular}{lccc}
\toprule
 & $\eta$ & $\nu$ & $\omega$ \\
\midrule
order $\partial^2$ & 0.04383(20) & 0.62841(9) & 0.84807(3) \\
order $\partial^4$ & 0.03435(47) & 0.63038(15) & 0.8256(5) \\
conformal bootstrap \cite{Kos:2016ysd,Simmons-Duffin:2016wlq} & 0.036298 & 0.629971 & 0.82968 \\
\bottomrule
\end{tabular}
\caption{Critical exponents on the front at the two orders. The brackets give half of the range over the front.}
\label{tab:orders}
\end{table}

The dependence of these results on the threshold in (\ref{eq:window}) has been examined by repeating the selection
with $1/20$ and with $1/5$ in the place of $1/10$, on the candidates with $m$ in steps of $0.25$. The amplitude selected for a given sharpness changes by one percent
at most. The interval of the sharpness is more sensitive: at order $\partial^2$ it becomes $6.5\le m\le7.75$ and
$m\le8$, respectively, and at order $\partial^4$ it becomes $5.5\le m\le10.25$ and the whole range which has been
scanned. With the threshold $1/5$ the interval ends closer to the minimum of the potential, a part of the region
where the fluctuations are active is left out, and smoother regulators are then admitted. With the threshold $1/20$ the exponents on the front remain within the intervals (\ref{eq:exp2}) and
(\ref{eq:exp4}). With $1/5$ the smooth regulators which enter the front extend the intervals: at order $\partial^2$
to $\eta\le0.04446$, $\nu\ge0.62808$ and $\omega\le0.84828$, and at order $\partial^4$
to $\eta\ge0.03373$, $\nu\le0.63058$ and $\omega\ge0.8248$.

The exponential regulator is dominated at this order as well, by a factor of about $1.5$ (Fig.~\ref{fig:front}). The compact
regulator $r=a(1-y)^3\theta(1-y)$ of Ref.~\cite{Codello:2026xyz} has an amplitude and no sharpness, and it is
therefore not among the candidates of the selection. Its defects have nevertheless been evaluated. At the amplitude at
which $\eta$ is stationary, which is the regulator selected in that work, they are $(\Delta_V,\Delta_Z)=(0.035,0.0156)$, both
larger than those of the cyclotomic regulator at the smooth end of the front. For amplitudes between $0.2\%$ and $0.8\%$ below the
stationary one its $\Delta_Z$ is smaller than that of any cyclotomic regulator, by $7\%$ at most, with $\Delta_V$ nearly unchanged; this is not the
regulator selected by the stationarity of $\eta$. In the plane of Fig.~\ref{fig:front} it lies beyond the smooth end
of the front, with a defect in the potential sector which is between $1.5$ and $24$ times larger than on the front,
and its exponents are not included in (\ref{eq:exp4}). The regulator with three parameters optimized in
Ref.~\cite{Codello:2026xyz} could not be evaluated, since its coefficients are not given there.

\section{Relation with the principle of minimal sensitivity}
\label{sec:pms}

The amplitude $\alpha_s(m)$ at which $\eta$ is stationary can be computed, after the selection, from the same
fixed points. For the values of $m$ at which it has been computed, the regulators of the front lie between $0.989\,\alpha_s$ and $0.992\,\alpha_s$ at order
$\partial^2$ and between $0.997\,\alpha_s$ and $1.004\,\alpha_s$ at order $\partial^4$. The selected amplitude is thus
close to the PMS amplitude of $\eta$, although it does not coincide with it.

At large field the loop term in (\ref{eq:fp2}) or (\ref{eq:fp4}) is negligible and
\begin{equation}
v(x)\simeq A\,x^{\,d/D_\varphi}\,,\qquad \frac{d}{D_\varphi}=\frac{2d}{d-2+\eta}\,.
\label{eq:largefield}
\end{equation}
Both the amplitude $A$ and the exponent depend on the regulator, so that the response $\partial_\theta v^*$ behaves
as
\begin{equation}
\frac{\partial_\theta v^*}{v}\simeq\partial_\theta\ln A-\frac{2d\,\partial_\theta\eta}{(d-2+\eta)^2}\,\ln x\,.
\label{eq:logterm}
\end{equation}
On the other hand, in the same region the loop term is negligible and the response prescribed by the identity in
(\ref{eq:defectdef}) reduces to $\gamma N_v\simeq-\gamma\,(d/2D_\varphi)\,v$, so that the defect of the potential
behaves as
\begin{equation}
\frac{D_v}{v}\simeq\partial_\theta\ln A+\frac{d}{2D_\varphi}\,\gamma-\frac{2d\,\partial_\theta\eta}{(d-2+\eta)^2}\,\ln x\,.
\label{eq:defectlarge}
\end{equation}
The first two terms do not depend on the field and cancel for a suitable value of $\gamma$, while the logarithm cannot
be removed by any choice of $\gamma$. A defect which vanishes for all $x\geq0$ is therefore possible only if
\begin{equation}
\partial_\theta\,\eta=0\,.
\label{eq:pmseta}
\end{equation}
The stationarity of $\eta$ with respect to the amplitude is thus a necessary condition for a defect which vanishes for
all values of the field. The argument extends to order $\partial^4$ if the fixed point has the same behavior at large field and
the loop terms are subleading there. The minima found on the interval (\ref{eq:window}) are different from zero, and
their proximity to the stationary amplitude is a numerical result, consistent with this condition.

The condition (\ref{eq:pmseta}) is necessary and not sufficient. The regulators in common use of Table~\ref{tab:standard} satisfy it
by construction and are nevertheless far from the front; the stationarity of $\eta$ does not select the sharpness
either, since along the line $\alpha=\alpha_s(m)$ the anomalous dimension has its own stationary point at a sharpness
which is outside the front at order $\partial^2$.

No statement similar to (\ref{eq:pmseta}) holds for the other exponents. The amplitudes at which $\nu$ and $\omega$
are stationary are given in Table~\ref{tab:pms} in units of $\alpha_s$. At order $\partial^2$ the exponent $\nu$ has a
maximum $3$--$4\%$ below $\alpha_s$; at order $\partial^4$ it has a minimum $2$--$5\%$ above it. The exponent
$\omega$ has a minimum close to $\alpha_s$ at order $\partial^2$, while at order $\partial^4$ it has no stationary
point in the neighbourhood of $\alpha_s$, the nearest one being a maximum at $1.4\,\alpha_s$. The identity selects
neither of these amplitudes. The procedure in which every exponent is evaluated at its own stationary point thus
makes use of three different regulators, only one of which is close to the one singled out by the identity, and for
$\omega$ such a point may not exist. The numerical consequences for $\nu$ are small, because
this exponent varies slowly with the amplitude: its value at the selected amplitude differs from its stationary value
by less than $2\times10^{-5}$, which is well inside the interval (\ref{eq:exp4}).

\begin{table}[t]
\centering
\begin{tabular}{lcccc}
\toprule
 & $m$ & $\nu$ stationary & $\omega$ stationary & $\Delta_Z$ minimal \\
\midrule
order $\partial^2$ & 5 & 0.963 & 1.011 & 0.989 \\
                   & 7 & 0.968 & 1.004 & 0.990 \\
                   & 8 & 0.969 & 0.994 & 0.990 \\
\midrule
order $\partial^4$ & 5 & 1.019 & -- & 1.002 \\
                   & 7 & 1.032 & 1.41 & 1.004 \\
                   & 10 & 1.043 & 1.39 & 1.005 \\
\bottomrule
\end{tabular}
\caption{Amplitudes at which $\nu$ and $\omega$ are stationary and amplitude at which $\Delta_Z$ is minimal, in units of
the amplitude $\alpha_s$ at which $\eta$ is stationary. A dash means that no stationary point is found between
$0.6\,\alpha_s$ and $1.4\,\alpha_s$.}
\label{tab:pms}
\end{table}

\section{Conclusions}
\label{sec:conclusions}

A criterion for the choice of the regulator in truncated functional RG flows has been presented, in which the
regulator is selected by the violation of the exact identity (\ref{eq:identity}) and no observable is used. For the
Ising fixed point in three dimensions, with the cyclotomic regulator (\ref{eq:cyclotomic}), the two relative defects
of the potential and of the wave-function sector fix the amplitude of the regulator and restrict its sharpness to an
interval, at order $\partial^2$ and at order $\partial^4$ of the derivative expansion. The critical exponents on the
selected set of regulators are given in Table~\ref{tab:orders}; at each regulator of the front the three exponents belong to the same fixed point, and
the values of the conformal bootstrap lie between those of the two orders.

The stationarity of the anomalous
dimension with respect to the amplitude follows from the identity as a necessary condition, and the selected amplitude
lies within about one percent of the stationary one. The stationarity of $\nu$ and of $\omega$ does not follow,
and the corresponding amplitudes are different. The regulators in common use, taken at the amplitude at which $\eta$
is stationary, violate the identity more than the selected ones in both sectors.

Since the construction relies only on the structure of the flow equation, it can be applied to problems in which the
principle of minimal sensitivity is not available because no suitable observable exists. The regulators which depend
on the fields, for which the identity acquires additional contributions \cite{Bonanno:2026bilocal}, are the natural
next step.

\section*{Acknowledgments}

The author is grateful to Renata Ferrero, Oleg Melichev and Dario Zappal\`a for a critical reading of the manuscript and for many
discussions. Large language models have been used as tools in this work: Claude (Anthropic) for the symbolic and
numerical programming and in the preparation of the manuscript, and Codex (OpenAI) for independent checks of the
derivations and of the numerical results. Access to Claude was provided through the Claude Team for Scientists
program of Anthropic. The ideas, the choice of the methods, the interpretation of the results
and the conclusions are the author's, who takes full responsibility for the content of the paper.

\appendix

\section{Flow equations at order $\partial^4$}
\label{app:flows}

The flow equations of the five functions of the ansatz (\ref{eq:ansatz4}) are collected here in the dimensionful
form, for a general dimension $d$ and a general regulator, in the notation of Sec.~\ref{sec:defect}. The propagator
at constant field is
\begin{equation}
G_k(q^2,\varphi)=\frac{1}{W^a_k\,q^4+Z_k\,q^2+V_k''+R_k(q^2)}\,,
\label{eq:appG}
\end{equation}
a prime on the functions of the ansatz denotes a derivative with respect to $\varphi$ and a prime on $G_k$ a
derivative with respect to $q^2$. The $Q$-functionals are those of Eq.~(\ref{eq:Qfun}). The vertices which
follow from (\ref{eq:ansatz4}) at constant field, with incoming momenta $p_1,\dots,p_n$, are
\begin{align}
\Gamma_k^{(n)}={}&V_k^{(n)}-Z_k^{(n-2)}\sum_{i<j}p_i\!\cdot\!p_j
+W_k^{a\,(n-2)}\sum_{i<j}(p_i\!\cdot\!p_j)^2
+W_k^{b\,(n-3)}\sum_{i<j}\,\sum_{l\neq i,j}(p_i\!\cdot\!p_j)\,p_l^2\nonumber\\
&+4\,W_k^{c\,(n-4)}\sum_{\{i,j\},\{l,m\}}(p_i\!\cdot\!p_j)(p_l\!\cdot\!p_m)\,,
\label{eq:vertices}
\end{align}
where a superscript in brackets is the order of the derivative with respect to $\varphi$, a function with a negative
order of derivative is absent, and the last sum runs over the unordered pairs of disjoint pairs of legs. The flow of
the $n$-point function is obtained from (\ref{eq:flow}) by taking $n$ derivatives with respect to the field. The
equation for $Z_k$ and $W^a_k$ follows from the terms of order $p^2$ and $p^4$ of the two-point function,
the one for $W^b_k$ from the three-point function and the one for $W^c_k$ from the four-point function, after the
expansion in the external momenta and the angular average. The flow equations of this order have also been given in
Ref.~\cite{Codello:2026xyz}. The expressions used here were derived independently; they are given in full in the
ancillary files and have been verified term by term for $Z_k$ and $W^a_k$, and numerically for $W^b_k$ and $W^c_k$. The potential obeys
Eq.~(\ref{eq:BV}) with the propagator (\ref{eq:appG}), and for the two functions of the two-point sector one finds
\begin{align}
(4\pi)^{\frac d2}\,\partial_tZ_k={}&\big(V_k'''\big)^{2}\,\Big\{Q_{\frac d2}\big[G_k^{2}G_k'\,\partial_tR_k\big] + Q_{\frac d2+1}\big[G_k^{2}G_k''\,\partial_tR_k\big]\Big\}\nonumber\\
&+V_k'''\,Z_k'\,\Big\{2\,Q_{\frac d2}\big[G_k^{3}\,\partial_tR_k\big] + (d + 2)\,Q_{\frac d2+1}\big[G_k^{2}G_k'\,\partial_tR_k\big]\nonumber\\
&\qquad + (d + 2)\,Q_{\frac d2+2}\big[G_k^{2}G_k''\,\partial_tR_k\big]\Big\}\nonumber\\
&+V_k'''\,{W^a_k}'\,\Big\{3\,Q_{\frac d2+1}\big[G_k^{3}\,\partial_tR_k\big]\nonumber\\
&\qquad + \frac{\left(d + 2\right) \left(d + 4\right)}{2}\,Q_{\frac d2+2}\big[G_k^{2}G_k'\,\partial_tR_k\big]\nonumber\\
&\qquad + \frac{\left(d + 2\right) \left(d + 4\right)}{2}\,Q_{\frac d2+3}\big[G_k^{2}G_k''\,\partial_tR_k\big]\Big\}\nonumber\\
&-2 \left(d - 1\right)\,V_k'''\,{W^b_k}\,Q_{\frac d2+1}\big[G_k^{3}\,\partial_tR_k\big]\nonumber\\
&+\big(Z_k'\big)^{2}\,\Big\{\frac{2 d + 1}{2}\,Q_{\frac d2+1}\big[G_k^{3}\,\partial_tR_k\big]\nonumber\\
&\qquad + \frac{\left(d + 2\right) \left(d + 4\right)}{4}\,Q_{\frac d2+2}\big[G_k^{2}G_k'\,\partial_tR_k\big]\nonumber\\
&\qquad + \frac{\left(d + 2\right) \left(d + 4\right)}{4}\,Q_{\frac d2+3}\big[G_k^{2}G_k''\,\partial_tR_k\big]\Big\}\nonumber\\
&+{W^a_k}'\,Z_k'\,\Big\{\frac{\left(d + 2\right) \left(d + 5\right)}{2}\,Q_{\frac d2+2}\big[G_k^{3}\,\partial_tR_k\big]\nonumber\\
&\qquad + \frac{\left(d + 2\right) \left(d + 4\right) \left(d + 6\right)}{4}\,Q_{\frac d2+3}\big[G_k^{2}G_k'\,\partial_tR_k\big]\nonumber\\
&\qquad + \frac{\left(d + 2\right) \left(d + 4\right) \left(d + 6\right)}{4}\,Q_{\frac d2+4}\big[G_k^{2}G_k''\,\partial_tR_k\big]\Big\}\nonumber\\
&-\left(d - 1\right) \left(d + 2\right)\,{W^b_k}\,Z_k'\,Q_{\frac d2+2}\big[G_k^{3}\,\partial_tR_k\big]\nonumber\\
&-\frac{1}{2}\,Z_k''\,Q_{\frac d2}\big[G_k^{2}\,\partial_tR_k\big]\nonumber\\
&+\big({W^a_k}'\big)^{2}\,\Big\{\frac{5 \left(d + 2\right) \left(d + 4\right)}{4}\,Q_{\frac d2+3}\big[G_k^{3}\,\partial_tR_k\big]\nonumber\\
&\qquad + \frac{\left(d + 2\right) \left(d + 4\right) \left(d + 6\right) \left(d + 8\right)}{16}\,Q_{\frac d2+4}\big[G_k^{2}G_k'\,\partial_tR_k\big]\nonumber\\
&\qquad + \frac{\left(d + 2\right) \left(d + 4\right) \left(d + 6\right) \left(d + 8\right)}{16}\,Q_{\frac d2+5}\big[G_k^{2}G_k''\,\partial_tR_k\big]\Big\}\nonumber\\
&-\frac{\left(d - 1\right) \left(d + 2\right) \left(d + 4\right)}{2}\,{W^a_k}'\,{W^b_k}\,Q_{\frac d2+3}\big[G_k^{3}\,\partial_tR_k\big]\nonumber\\
&-{W^a_k}''\,Q_{\frac d2+1}\big[G_k^{2}\,\partial_tR_k\big]\nonumber\\
&+d\,{W^b_k}'\,Q_{\frac d2+1}\big[G_k^{2}\,\partial_tR_k\big]\nonumber\\
&-(d + 2)\,{W^c_k}\,Q_{\frac d2+1}\big[G_k^{2}\,\partial_tR_k\big]\,.
\label{eq:flowZ}
\end{align}
\begin{align}
(4\pi)^{\frac d2}\,\partial_tW^a_k={}&\big(V_k'''\big)^{2}\,\Big\{\frac{1}{2}\,Q_{\frac d2}\big[G_k^{2}G_k''\,\partial_tR_k\big] + Q_{\frac d2+1}\big[G_k^{2}G_k'''\,\partial_tR_k\big]\nonumber\\
&\qquad + \frac{1}{2}\,Q_{\frac d2+2}\big[G_k^{2}G_k''''\,\partial_tR_k\big]\Big\}\nonumber\\
&+V_k'''\,Z_k'\,\Big\{2\,Q_{\frac d2}\big[G_k^{2}G_k'\,\partial_tR_k\big] + \frac{d + 8}{2}\,Q_{\frac d2+1}\big[G_k^{2}G_k''\,\partial_tR_k\big]\nonumber\\
&\qquad + (d + 4)\,Q_{\frac d2+2}\big[G_k^{2}G_k'''\,\partial_tR_k\big] + \frac{d + 4}{2}\,Q_{\frac d2+3}\big[G_k^{2}G_k''''\,\partial_tR_k\big]\Big\}\nonumber\\
&+V_k'''\,{W^a_k}'\,\Big\{2\,Q_{\frac d2}\big[G_k^{3}\,\partial_tR_k\big]\nonumber\\
&\qquad + 7\,Q_{\frac d2+1}\big[G_k^{2}G_k'\,\partial_tR_k\big]\nonumber\\
&\qquad + \frac{d^{2} + 10 d + 52}{4}\,Q_{\frac d2+2}\big[G_k^{2}G_k''\,\partial_tR_k\big]\nonumber\\
&\qquad + \frac{\left(d + 4\right) \left(d + 6\right)}{2}\,Q_{\frac d2+3}\big[G_k^{2}G_k'''\,\partial_tR_k\big]\nonumber\\
&\qquad + \frac{\left(d + 4\right) \left(d + 6\right)}{4}\,Q_{\frac d2+4}\big[G_k^{2}G_k''''\,\partial_tR_k\big]\Big\}\nonumber\\
&+V_k'''\,{W^b_k}\,\Big\{-2 \left(d - 1\right)\,Q_{\frac d2+1}\big[G_k^{2}G_k'\,\partial_tR_k\big] - 2 \left(d - 1\right)\,Q_{\frac d2+2}\big[G_k^{2}G_k''\,\partial_tR_k\big]\Big\}\nonumber\\
&+\big(Z_k'\big)^{2}\,\Big\{Q_{\frac d2}\big[G_k^{3}\,\partial_tR_k\big]\nonumber\\
&\qquad + \frac{2 d + 5}{2}\,Q_{\frac d2+1}\big[G_k^{2}G_k'\,\partial_tR_k\big]\nonumber\\
&\qquad + \frac{d^{2} + 18 d + 44}{8}\,Q_{\frac d2+2}\big[G_k^{2}G_k''\,\partial_tR_k\big]\nonumber\\
&\qquad + \frac{\left(d + 4\right) \left(d + 6\right)}{4}\,Q_{\frac d2+3}\big[G_k^{2}G_k'''\,\partial_tR_k\big]\nonumber\\
&\qquad + \frac{\left(d + 4\right) \left(d + 6\right)}{8}\,Q_{\frac d2+4}\big[G_k^{2}G_k''''\,\partial_tR_k\big]\Big\}\nonumber\\
&+{W^a_k}'\,Z_k'\,\Big\{(d + 5)\,Q_{\frac d2+1}\big[G_k^{3}\,\partial_tR_k\big]\nonumber\\
&\qquad + \frac{\left(d + 4\right) \left(d + 11\right)}{2}\,Q_{\frac d2+2}\big[G_k^{2}G_k'\,\partial_tR_k\big]\nonumber\\
&\qquad + \frac{\left(d + 4\right) \left(d^{2} + 18 d + 92\right)}{8}\,Q_{\frac d2+3}\big[G_k^{2}G_k''\,\partial_tR_k\big]\nonumber\\
&\qquad + \frac{\left(d + 4\right) \left(d + 6\right) \left(d + 8\right)}{4}\,Q_{\frac d2+4}\big[G_k^{2}G_k'''\,\partial_tR_k\big]\nonumber\\
&\qquad + \frac{\left(d + 4\right) \left(d + 6\right) \left(d + 8\right)}{8}\,Q_{\frac d2+5}\big[G_k^{2}G_k''''\,\partial_tR_k\big]\Big\}\nonumber\\
&+{W^b_k}\,Z_k'\,\Big\{-2 \left(d - 1\right)\,Q_{\frac d2+1}\big[G_k^{3}\,\partial_tR_k\big]\nonumber\\
&\qquad - \left(d - 1\right) \left(d + 4\right)\,Q_{\frac d2+2}\big[G_k^{2}G_k'\,\partial_tR_k\big]\nonumber\\
&\qquad - \left(d - 1\right) \left(d + 4\right)\,Q_{\frac d2+3}\big[G_k^{2}G_k''\,\partial_tR_k\big]\Big\}\nonumber\\
&+\big({W^a_k}'\big)^{2}\,\Big\{\frac{2 d^{2} + 12 d + 43}{4}\,Q_{\frac d2+2}\big[G_k^{3}\,\partial_tR_k\big]\nonumber\\
&\qquad + \frac{9 \left(d + 4\right) \left(d + 6\right)}{4}\,Q_{\frac d2+3}\big[G_k^{2}G_k'\,\partial_tR_k\big]\nonumber\\
&\qquad + \frac{\left(d + 4\right) \left(d + 6\right) \left(d^{2} + 18 d + 152\right)}{32}\,Q_{\frac d2+4}\big[G_k^{2}G_k''\,\partial_tR_k\big]\nonumber\\
&\qquad + \frac{\left(d + 4\right) \left(d + 6\right) \left(d + 8\right) \left(d + 10\right)}{16}\,Q_{\frac d2+5}\big[G_k^{2}G_k'''\,\partial_tR_k\big]\nonumber\\
&\qquad + \frac{\left(d + 4\right) \left(d + 6\right) \left(d + 8\right) \left(d + 10\right)}{32}\,Q_{\frac d2+6}\big[G_k^{2}G_k''''\,\partial_tR_k\big]\Big\}\nonumber\\
&+{W^a_k}'\,{W^b_k}\,\Big\{-3 \left(d - 1\right)\,Q_{\frac d2+2}\big[G_k^{3}\,\partial_tR_k\big]\nonumber\\
&\qquad - \frac{\left(d - 1\right) \left(d + 4\right) \left(d + 6\right)}{2}\,Q_{\frac d2+3}\big[G_k^{2}G_k'\,\partial_tR_k\big]\nonumber\\
&\qquad - \frac{\left(d - 1\right) \left(d + 4\right) \left(d + 6\right)}{2}\,Q_{\frac d2+4}\big[G_k^{2}G_k''\,\partial_tR_k\big]\Big\}\nonumber\\
&-\frac{1}{2}\,{W^a_k}''\,Q_{\frac d2}\big[G_k^{2}\,\partial_tR_k\big]\nonumber\\
&+\left(d - 1\right) \left(d + 1\right)\,\big({W^b_k}\big)^{2}\,Q_{\frac d2+2}\big[G_k^{3}\,\partial_tR_k\big]\,.
\label{eq:flowWa}
\end{align}

For $W^a_k=W^b_k=W^c_k=0$ Eq.~(\ref{eq:flowZ}) reduces to (\ref{eq:BZ}). The equations for $W^b_k$ and $W^c_k$ have the same
structure and contain $182$ and $561$ terms, respectively; they are given in the ancillary file
\texttt{flow\_equations\_d4.txt}, together with (\ref{eq:flowZ}) and (\ref{eq:flowWa}) in the same format, and the four
equations are typeset in the ancillary file \texttt{flow\_equations\_d4\_typeset.pdf}. The
three-point and the four-point functions also determine the first and the second derivative of the right-hand side
of (\ref{eq:flowWa}) with respect to the field, and the first derivative of the equation for $W^b_k$; these relations
have been used as a check and are satisfied identically. The equations (\ref{eq:fp4}) follow with the rescaling
(\ref{eq:rescaling}), extended to $W^a_k=\bar Z_k\,k^{-2}\,w_a$,
$W^b_k=\bar Z_k^{3/2}\,k^{-(d+2)/2}\,w_b$ and $W^c_k=\bar Z_k^{2}\,k^{-d}\,w_c$, and with the factor
$(4\pi)^{-d/2}$ absorbed in a further rescaling of $x$ and of the five functions, as at order $\partial^2$.

\section{Numerical methods}
\label{app:numerics}

At order $\partial^2$ the functions $u=v'/x$ and $z$ are computed as functions of $\rho=x^2/2$. They are represented
on $\rho\in[0,\rho_m]$ by a Chebyshev expansion and on $[\rho_m,\infty)$ by an expansion in rational Chebyshev
functions of $u$ and $z$ divided by their asymptotic powers, $\rho^{(2-\eta)/(1+\eta)}$ and $\rho^{-\eta/(1+\eta)}$ in
$d=3$; the equations are imposed by collocation and the two domains are matched at $\rho_m$
\cite{Borchardt:2015rxa}. With polynomials of order $64$ and $48$ in the two domains and $480$
nodes in the momentum integrals $\eta$ is stable to twelve digits. The response $\partial_\theta f^*$ is obtained from
the linearized equations and has been compared with the difference of two fixed points computed with slightly
different regulators, with agreement to one part in $10^{8}$. For a variation which corresponds to a change of $k$
the defect is $10^{-11}$ of the response. The exponents $\nu$ and $\omega$ are obtained from the eigenvalues
$\lambda$ of the flow linearized around the fixed point, $\nu=-1/\lambda$ for the relevant one and $\omega=\lambda$
for the first irrelevant one, with the normalization condition (\ref{eq:norm}) imposed at the minimum
of the perturbed potential.

At order $\partial^4$ the five functions are represented on a uniform grid in $x$ with $140$ nodes up to
$x\simeq4.2\,x_0$, the derivatives are computed with stencils of thirteen points, continued by parity at the
origin, and no condition is imposed at the upper end. The momentum integrals are computed by Gauss-Legendre quadrature
up to $y=60$. Finer grids and stencils of different order change $\eta$, $\nu$ and $\omega$ by less than $10^{-10}$, and
an independent calculation with Chebyshev collocation and with the large-field behaviour imposed as a boundary
condition reproduces $\eta$ to the same accuracy; the function $w_c$ is the least accurate, with a relative
difference between the two calculations of a few times $10^{-4}$ in the interval (\ref{eq:window}) for the cyclotomic
regulator. For a variation
which corresponds to a change of $k$ the defect is $10^{-9}$ of the response for $m=7$ and $10^{-11}$ at the sharp
end of the front; at the smooth end, where the cut of the momentum integrals is felt most, it is $3\times10^{-5}$.

The fixed points of (\ref{eq:norm}) are mapped exactly onto those of the alternative normalization $z(0)=1$ by
$\alpha\to\alpha/z(0)$ and $x\to z(0)^{1/2}\,x$, the anomalous dimension being unchanged. The fronts of the two normalizations are not related by this map, since the response to a change of the amplitude is not the same in the two
normalizations (Appendix~\ref{app:norm}). A selection carried out in that normalization on the candidates with $m$ in steps of $0.25$, with
the amplitudes placed around the stationary one, gave the same interval of the sharpness and the same interval of
$\eta$ as the selection performed with (\ref{eq:norm}) on the same candidates; the exponents $\nu$ and
$\omega$ at a given fixed point differ in the fifth and in the fourth digit, respectively, since $Z_k$ enters the
regulator away from the fixed point.

\section{Closed forms for the cyclotomic regulator}
\label{app:closed}

\begin{figure}[t]
\centering
\includegraphics[width=0.45\textwidth]{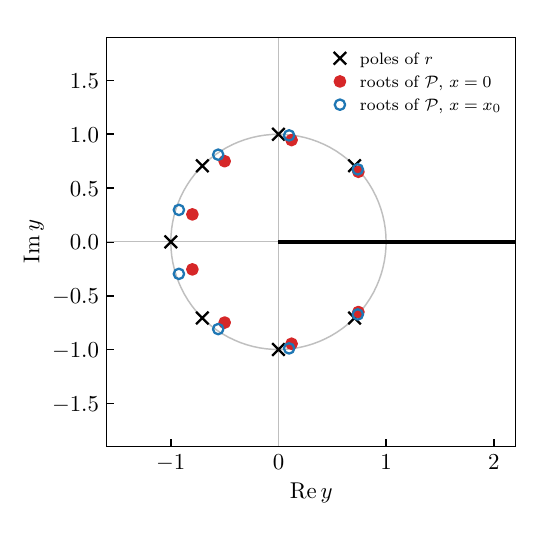}
\caption{Poles in the complex plane of $y$ for the cyclotomic regulator with $m=7$ and $\alpha=0.6055$ at order
$\partial^2$: the poles of $r$ (crosses), which lie on the unit circle, and the roots of the polynomial
(\ref{eq:Qpoly}), which are the poles of the propagator, at the origin and at the minimum of the potential of the
fixed point. The thick line is the integration path.}
\label{fig:poles}
\end{figure}

This appendix illustrates Eq.~(\ref{eq:residues}) in $d=3$. With the insertion (\ref{eq:rdot}) and $g=E/\mathcal P$
one has
\begin{equation}
g\,\dot r=\alpha\,\frac{(2-\eta)\,E+2\,y\,E'}{E\,\mathcal P}\,,\qquad
g^2\,\dot r=\alpha\,\frac{(2-\eta)\,E+2\,y\,E'}{\mathcal P^2}\,.
\label{eq:grdot}
\end{equation}
The first expression enters the loop term of the potential, $b_V[\dot r]=\pi^{-1/2}\int_0^\infty dy\,\sqrt y\,g\,\dot r$.
It has simple poles at the roots $y_j$ of $\mathcal P$ and at the roots of unity $\zeta\neq1$, the zeros of $E$. At the
latter $\mathcal P=\alpha$, the residue is $2\,\zeta$ whatever the fixed point, and their sum gives a number,
\begin{equation}
\int_0^\infty dy\,\sqrt y\;g\,\dot r=-2\pi\cot\frac{3\pi}{2\,(m+1)}
-\pi\,\alpha\sum_{j=1}^{m+1}\left.\frac{\big[(2-\eta)\,E+2\,y\,E'\big]\sqrt{-y}}{E\;\mathcal P'}\right|_{y=y_j}\,.
\label{eq:bVclosed}
\end{equation}
The first term does not depend on the field and drops from the equation for $u$. Accordingly, in the second
expression of (\ref{eq:grdot}), which is the one generated by a derivative with respect to the field, the polynomial
$E$ has cancelled, and only the roots of $\mathcal P$ are left; the same happens in all the terms of (\ref{eq:BZ}).
The poles are displayed in Fig.~\ref{fig:poles} for a regulator with $m=7$ close to the front. The roots of
$\mathcal P$ lie close to the unit circle and away from the positive real axis, along which the integration runs.

For $m=1$, $r=\alpha/(1+y)$, the sums can be performed without solving for the roots. Here
$\mathcal P=z\,(y+s_+^2)(y+s_-^2)$, and the results depend on $s_\pm$ only through
\begin{equation}
q=s_+s_-=\sqrt{\frac{v''+\alpha}{z}}\,,\qquad p=s_++s_-=\sqrt{1+\frac{v''}{z}+2\,q}\,.
\label{eq:pq}
\end{equation}
From the elementary integral $\int_0^\infty dy\,\sqrt y\,\prod_{i=1}^{3}(y+s_i^2)^{-1}
=\pi\,[(s_1+s_2)(s_2+s_3)(s_3+s_1)]^{-1}$ one finds
\begin{equation}
b_V[\dot r]=\frac{\sqrt\pi\,\alpha}{z}\;\frac{2-\eta+(4-\eta)\,(p+q)}{p\,(1+p+q)}\,.
\label{eq:bVm1}
\end{equation}
In the local potential approximation, $z=1$ and $\eta=0$, the fixed-point equation is then the ordinary differential
equation
\begin{equation}
0=-3\,v+\frac12\,x\,v'+2\sqrt\pi\,\alpha\;\frac{1+2\,p+2\,q}{p\,(1+p+q)}\,,\qquad
q=\sqrt{v''+\alpha}\,,\quad p=\sqrt{1+v''+2\,q}\,.
\label{eq:lpam1}
\end{equation}
The loop term (\ref{eq:loopterm}) of the equation for $u$ contains double poles only,
$\ell=-\pi^{-1/2}\int_0^\infty dy\,\sqrt y\,g^2\,\dot r\,(z'\,y+v''')/x$, and it reads
\begin{equation}
\ell=-\frac{\sqrt\pi\,\alpha}{2\,z^2\,p^3}\,\bigg\{\Big[\frac{2-\eta}{q}+4-\eta\Big]\frac{v'''}{x}
+\Big[2-\eta+(4-\eta)\,\big(p^2+q\big)\Big]\frac{z'}{x}\bigg\}\,.
\label{eq:ellm1}
\end{equation}
For $m=2$ and $m=3$ the roots are those of a cubic and of a quartic; for larger $m$ they are computed numerically.
The expressions of this appendix have been compared with the numerical quadrature, with which they agree to
$10^{-13}$.

\section{The number $\gamma$ and the normalization condition}
\label{app:norm}

Let
\begin{equation}
A_f\equiv\partial_\theta f^*-b_f[\delta r]\,,\qquad A_u=\frac{A_v'}{x}\,,
\label{eq:Adef}
\end{equation}
be the difference of the two responses without the term proportional to $N_f$, so that the defect
(\ref{eq:defectdef}) is $D_f=A_f-\gamma\,N_f$. The quantity $\Delta_V^2+\Delta_Z^2$ is quadratic in $\gamma$, and
its minimum is at
\begin{equation}
\gamma=\frac{\langle A_u,N_u\rangle\,\lVert\partial_\theta u^*\rVert^{-2}+\langle A_z,N_z\rangle\,\lVert\partial_\theta z^*\rVert^{-2}}
{\lVert N_u\rVert^{2}\,\lVert\partial_\theta u^*\rVert^{-2}+\lVert N_z\rVert^{2}\,\lVert\partial_\theta z^*\rVert^{-2}}\,,
\label{eq:gammafix}
\end{equation}
which removes from $A$ its component along $N$. This is the value used in the paper.

A different value of $\gamma$ is obtained if the normalization of the field is imposed on the response prescribed
by the identity. The fixed points of a family of regulators are computed with the normalization (\ref{eq:norm}) for every value of
$\theta$. Differentiating the two conditions with respect to $\theta$ and eliminating the displacement of the
minimum, the response of the truncated fixed point obeys
\begin{equation}
\partial_\theta z^*(x_0)-\frac{z'(x_0)}{v''(x_0)}\,\partial_\theta v^{*\prime}(x_0)=0\,.
\label{eq:linnorm}
\end{equation}
If the same condition is imposed on the response $b_f[\delta r]+\gamma\,N_f$ prescribed by the identity, a value
$\gamma_0$ is obtained,
\begin{equation}
\gamma_0=-\left.\frac{b_Z[\delta r]-\dfrac{z'}{v''}\,\big(b_V[\delta r]\big)'}{N_z-\dfrac{z'}{v''}\,N_v'}\right|_{x=x_0}\,,
\label{eq:gamma0}
\end{equation}
and with it the unprojected defect $D^{(0)}_f=A_f-\gamma_0\,N_f$, for which the combination (\ref{eq:linnorm}) vanishes at the
minimum. If the field is normalized at the origin, $z(0)=1$, the point does not move with $\theta$, the condition is
$b_Z[\delta r](0)+\gamma\,N_z(0)=0$, and
\begin{equation}
\gamma_0=-\frac{b_Z[\delta r](0)}{N_z(0)}\,,
\end{equation}
so that the unprojected defect of $z$ vanishes at the origin. The two cases are related exactly. Let $A$ denote the
normalization (\ref{eq:norm}) and $B$ another one, with $\bar Z_k^B=c\,\bar Z_k^A$; for $z(0)=1$ one has $c=z_A(0)$.
The same regulators and the same fixed points are described in $B$ by
\begin{equation}
r_B=\frac{r_A}{c}\,,\qquad x_B=c^{1/2}\,x_A\,,\qquad v_B(x_B)=v_A(x_A)\,,\qquad z_B(x_B)=\frac{z_A(x_A)}{c}\,,
\end{equation}
and, since $c$ depends on the regulator, by the parameter $\theta_B=\theta_A-\ln c$. With
$q\equiv d\ln c/d\theta_A$ and $J\equiv d\theta_A/d\theta_B=(1-q)^{-1}$ the unprojected defects obtained from the two
normalization conditions, with the respective values $\gamma_0^A$ and $\gamma_0^B$, obey
\begin{equation}
D^{(0)B}_f=J\,S\,\Big\{D^{(0)A}_f-\kappa_{AB}\,N^A_f\Big\}\,,\qquad \kappa_{AB}=\frac{\gamma_0^B}{J}-\gamma_0^A-q\,,
\label{eq:DAB}
\end{equation}
where $S$ is the rescaling of the field and of the functions given above, and $\kappa_{AB}$ is the same number in
all the sectors. Up to a relabelling, a change of the point at which the field is normalized thus adds to the unprojected
defect a multiple of $N$, and to the response of the fixed point a multiple of $f+\frac12\,x\,f'$ in the two sectors used for the
selection, $f=u,z$. The first effect is not
small: for $m=8$ at order $\partial^2$ the defect of the potential obtained with $\gamma_0$ is $0.002$ of the
response, and the term $\kappa_{AB}N_u$ which is added to it by the normalization at the origin is $0.03$. For this
reason the component of the defect along $N$ is not used, and $\gamma$ is determined by (\ref{eq:gammafix}). The
second effect remains. The norms of the responses set both the weights of the two sectors in (\ref{eq:gammafix})
and the denominators of the relative defects (\ref{eq:objectives}), so that the removal of the component along $N$
does not make $\Delta_V$ and $\Delta_Z$ independent of the normalization: between
the two normalizations $\Delta_V$ changes by about one percent and $\Delta_Z$ by about ten percent at order
$\partial^2$, with the consequences described in Appendix~\ref{app:numerics}.

\bibliographystyle{apsrev4-2}
\bibliography{refs}

\end{document}